\documentclass[12pt]{article}

\usepackage{epsfig} \usepackage{amsfonts, amssymb, amsthm, graphicx}

\usepackage{amsmath}

\usepackage{amsbsy}

\usepackage{accents}
\usepackage{tensor}

\newtheorem{de}{Definition}

\newtheorem{co}{Corollary}

\renewcommand{\theequation}{\thesection.\arabic{equation}}

\newcommand{\be}{\begin{equation}}

\newcommand{\ee}{\end{equation}}

\newcommand{\bea}{\begin{eqnarray}}

\newcommand{\eea}{\end{eqnarray}}

\newcounter{orange} \renewcommand{\theorange}{\alph{orange}}

\begin{document}

\title{\bf Formal Series  of Generalised Functions and Their Application to Deformation Quantisation} \author{ J. Tosiek \thanks{e.mail: tosiek@p.lodz.pl}  and M. Dobrski \thanks{e.mail: michal.dobrski@p.lodz.pl}\\
 {\em
Institute of Physics}\\ {\em \L\'{o}d\'{z} University of Technology}\\ {\em 90-924 \L\'{o}d\'{z}, Poland}}
 \date{\today} 
 \maketitle 
 \begin{abstract}
Foundations of the formal series $*$ -- calculus in deformation quantisation are discussed. Several classes of continuous linear functionals over  algebras applied in classical and quantum physics are introduced. The notion of  nonnegativity in formal series calculus is proposed. Problems with defining quantum states over the set of  formal series are analysed. 

  \end{abstract}


\section{Introduction}

Theoretical physics is oriented towards prediction new phenomena. This process is based on logic. We start from a few well established experimental laws and then using  mathematical techniques  postulate potential effects. Efficiency of this procedure  is incredible. That is why proper understanding and exploring  abstract structures standing behind physical reality is so fundamental. 

Our paper is focused on a  possible realisation of axioms constituting the phase  space version of quantum mechanics. At first sight they are obvious -- observables are represented by some real functions on the phase space of the system and states can be identified with real nonnegative functionals over the algebra of observables. Unfortunately,  presence of the Planck constant and a noncommutative way of multiplying observables generate serious obstacles.

The standard procedure of building an algebra of observables in phase space formulation of quantum mechanics is based on a formal series calculus with respect to the Planck constant. Thus in this essay the role of formal series in deformation quantisation is discussed. 

Although our considerations are inspired by physics, one can look at them in a more abstract way. Namely they refer to a mathematical problem of building a functional calculus over some deformed noncommutative algebra of  functions. To emphasise this universal aspect of our analysis we denote the deformation parameter by $\lambda.$ In physical applications it is identified with the Planck constant $\hbar.$

The postulate of formulating quantum physics at a phase space was presented in four outstanding works \cite{WY31, WI32, GW46, MO49}. And in
three of them \cite{WI32, GW46, MO49}  the power expansion of functions with respect to the deformation parameter was used. A straightforward consequence of dealing with
series was substituting integral formulas by differential ones. This trend was continued, when Bayen and coworkers \cite{baf, bay}
 proposed a modern version of deformation quantisation. They introduced a $*$--product in the algebra of functions
$C^{\infty}({\cal M})$ in the form 
\[
 f*g= f \cdot g + \sum_{k=1}^{\infty}\lambda^k C_k(f,g) 
 \]
  with bidifferential operators $C_k.$
This formula has determined the direction of evolution of the mathematical theory of deformation of algebras. 
 
The idea of doing calculations on formal series simplifies them a lot but causes the huge problem -- it destroys convergence. Whenever we know that a finite number of initial terms sufficiently approximates an analysed quantity, there is no reason to worry. This is why perturbation methods are so efficient.    But if we try to include infinite series in our calculus,
 several difficulties in  performing a consistent construction of the space of quantum states come to light. 

The text is addressed mainly to physicists.  Hence discussing  mathematical structure of the presented model we often refer to physical circumstances. The structure of our paper is determined by a triple of objects: an algebra of smooth functions $C^{\infty}({\cal M})$ with pointwise multiplication, its trivial deformation $C^{\infty}[\lambda^{-1}, \lambda]]({\cal M})$ to the set of formal series with an Abelian $\bullet$ -- product and the quantum algebra of formal series $C^{\infty}[\lambda^{-1}, \lambda]]({\cal M})$ with a noncommutative $*$  -- product. One may doubt if the second ingredient  is really 
worth studying. However, in our opinion it enlightens very well troubles caused by operating on formal series. Moreover, it appears as an important component of the  generalised function calculus over the ring  $(C^{\infty}[\lambda^{-1}, \lambda]]({\cal M}),*).$ 

We avoid strict detailed proofs because the calculus presented in the current paper is based on the well known theory of generalised functions. 
\section{Space of formal series} 

At the beginning we recall a few widely accepted assumptions about the system under considerations. 

Firstly the stage, on which the system is described, is a phase space. The phase space is a real differentiable manifold ${\cal M}$. Thus the phase space is equipped
with a topology but there is no norm or metric on it. Of course ${\cal M}$ need not be a vector space. Moreover, a symplectic
structure $\omega$ has been introduced on the manifold ${\cal M}$. Then an integration on ${\cal M}$ is defined. The existence of the
symplectic structure implies that the dimension of ${\cal M}$ is even.

Secondly, we postulate that classical observables are smooth real functions on the manifold ${\cal M}.$ One can discuss if this
condition is too strong, but in fact most of measurable quantities belong to this set of functions. Thus observables form a subset of
the commutative ring $(C^{\infty}({\cal M}),+,\cdot)$ of complex valued smooth functions. Moreover, the ring $(C^{\infty}({\cal M}),+,\cdot)$ is an algebra over 
${\mathbb C}.$ The constant function equal to ${\bf 1}$ at every point of the manifold ${\cal M}$ is the identity element of
this algebra. To shorten notation we will denote the set of functions $C^{\infty}({\cal M}),$ the ring $(C^{\infty}({\cal M}),+,\cdot),$ the vector space $(C^{\infty}({\cal M}),{\mathbb C},+,\cdot)$ and the algebra by the same symbol $C^{\infty}({\cal M}).$

The convergence of  a sequence $\{f_n\}_{n=1}^{\infty}$ of elements of the space $C^{\infty}({\cal M})$ is defined in the following way
\cite{sch}. We say that the sequence $\{f_n\}_{n=1}^{\infty}$ is convergent to a function $f_0,$ if on every compact subset of the
manifold ${\cal M}$ every sequence of partial derivatives $\Big\{\frac{ \partial^{m_1+m_2+\ldots + m_{2r}}}{\partial^{m_1}q^1 \ldots
\partial^{m_{2r}}q^{2r} }f_n \Big\}_{n=1}^{\infty}$ is uniformly convergent to the derivative $\frac{ \partial^{m_1+m_2+\ldots +
m_{2r}}}{\partial^{m_1}q^1 \ldots \partial^{m_{2r}}q^{2r} }f_0.$ The symbols $q^1, \ldots, q^{2r} $ denote local coordinates on ${\cal M}$, where $\dim {\cal M}=2r.$ The proposed definition of convergence implies that the function $f_0$ is also a function belonging to the
space $C^{\infty}({\cal M})$ and thus this space is complete with respect to the proposed topology.

Assume that a theory is considered, in which a fundamental constant $\lambda$ appears. For example in quantum mechanics this constant
may be identified with the Planck constant $\hbar.$ It is natural that $\lambda$ should be incorporated in the observables. Let us try to build a counterpart of the algebra  $C^{\infty}({\cal M})$ for this case.

A formal series in $\lambda$ of smooth functions is every expression of the form 
\be
 \label{1} 
 \varphi[[\lambda]]:= \sum_{l=0}^{\infty}\lambda^l
\varphi_l\;\;, \;\; \forall \;l \;\; \frac{\partial \varphi_l}{\partial \lambda}=0 \;\;{\rm and } \;\; \varphi_l \in C^{\infty}({\cal
M}).
 \ee 
 It is assumed that $\lambda \in {\mathbb R}$ and is positive. 
 Reality and positivity of $\lambda$ are required in order  to introduce real nonnegative functionals.
 We usually do not refer to any fixed value of $\lambda.$ Powers of this parameter are rather used to distinguish between the functions $\varphi_l.$  Since the series $\varphi[[\lambda]]$ is formal, we do not
consider its convergence.
The lack of convergence is frustrating from the mathematical point of view. But technical advantages of formal series calculus in many areas of physics are undisputed.

The idea of formal series arose from the Taylor expansion of smooth functions with respect to the parameter $\lambda$ around the point
$\lambda=0.$ Indeed, if a function $f$ depends on $\lambda$ in a way, in which the mentioned Taylor expansion exists \[ f \sim
\sum_{l=0}^{\infty}\lambda^l f_l\;\; , \;\; f_l:= \frac{1}{l!} \frac{\partial^{\,l} f}{\partial \lambda^l} \Big|_{\lambda=0}, \] one
might try to represent the function $f$ by its power expansion with respect to $\lambda.$ Of course there is no one-to-one
relationship between the function and its expansion. However, especially if the parameter $\lambda$ is small, one can treat a finite
part of this Taylor expansion as some reasonable approximation of the  function $f$.

Let us build a calculus of the formal series (\ref{1}). The set of formal series  with the natural addition
 \be
 \label{a1}
\sum_{l=0}^{\infty}\lambda^l \varphi_l + \sum_{k=0}^{\infty}\lambda^k \psi_k:= \sum_{l=0}^{\infty}\lambda^l (\varphi_l + \psi_l)
 \ee
constitutes an Abelian group. One can observe that this collection of formal series, denoted as $C^{\infty}[[\lambda]]({\cal M}),$ may be identified with the Cartesian product
\[ 
C^{\infty}[[\lambda]]({\cal M}) := {\mbox {\Large {$\times$}}}_{l=0}^{\infty}
C^{\infty}({\cal M})_l\;, \;\;\;\; \forall\; l\;\;\;C^{\infty}({\cal M})_l := C^{\infty}({\cal M}). \]

\subsection{Field of formal series of complex numbers}

To keep consistency of the constructed calculus for the formal series, we need to extend the field of complex numbers to some field depending on  the deformation parameter $\lambda.$ The best candidate seems to be the set
${\mathbb C} [[\lambda]]$ defined as
\be 
\label{2} 
{\mathbb C} [[\lambda]]:= {\mbox {\Large {$\times$}}}_{l=0}^{\infty} {\mathbb
C}_l, \;\;\;\; \forall\; l\;\;\;{\mathbb C}_l:={\mathbb C}
\ee 
with elements of the form
\[ 
{\mathbb C} [[\lambda]] \ni c[[\lambda]] = \sum_{l=0}^{\infty}
\lambda^l c_l, \;\;\;\; \forall\; l\; \;\;c_l \in {\mathbb C}.
 \] 
 The summation and the multiplication in ${\mathbb C} [[\lambda]]$
are introduced by the following rules
\be
\label{a2} 
\sum_{l=0}^{\infty} \lambda^l c_l + \sum_{k=0}^{\infty} \lambda^k d_k := \sum_{l=0}^{\infty} \lambda^l (c_l + d_l)
\;\;,\;\; \sum_{l=0}^{\infty} \lambda^l c_l \cdot \sum_{k=0}^{\infty} \lambda^k d_k := \sum_{z=0}^{\infty} \lambda^z \sum_{l=0}^{z}
c_l d_{z-l}.
 \ee
 However, the ring $({\mathbb C} [[\lambda]],+, \cdot )$ {\bf is not} a field. Indeed, any element $\sum_{l=0}^{\infty}\lambda^l c_l$ with $c_0=0$ is not reversible. Thus to turn this ring into a field we must expand it to a set ${\mathbb C} [\lambda^{-1},\lambda]]$ with expressions containing finite number of terms standing at negative powers of $\lambda.$ In general elements of  this set are of the form
 \be
 \label{uzup1}
 {\mathbb C} [\lambda^{-1}, \lambda]] \ni c[[\lambda]] = \sum_{l=-u}^{\infty}
\lambda^l c_l, \;\;\;\; \forall\; l\; \;\;c_l \in {\mathbb C}, \;\;\; u \in {\mathcal N}.
 \ee
We will apply reversibility in order to perform normalisation of functionals.

 Structure of the set ${\mathbb C} [\lambda^{-1}, \lambda]]$ is still given by (\ref{2}).
 The triad $({\mathbb C} [\lambda^{-1}, \lambda]],+, \cdot )$, where the operations `$+$' and `$\cdot$' have been introduced as natural generalisations of (\ref{a2}), constitutes a field. Notice that for some elements the limit $\lim_{\lambda \rightarrow 0^+} \,c[[\lambda]]$   may not exist. The field $({\mathbb C} [\lambda^{-1}, \lambda]],+, \cdot )$ will be simply denoted as  $({\mathbb C} [\lambda^{-1}, \lambda]].$

A sequence $\{(\sum_{l=-u}^{\infty} \lambda^l c_l)_n \}_{n=1}^{\infty}$ of elements from the field ${\mathbb C}
[\lambda^{-1},\lambda]]$ is convergent to an element $\sum_{l=-u}^{\infty} \lambda^l c_{l0}$ from ${\mathbb C} [\lambda^{-1},\lambda]],$ if for every index $l$
the series $\{( c_l)_n \}_{n=1}^{\infty}$ of complex numbers approaches $c_{l0}.$ Thus e.g. 
 $\{\sum_{l=0}^{\infty} \lambda^l \delta_{l n}\}_{n=1}^{\infty} \rightarrow 0,$ where by $ \delta_{l n}$ we mean the Kronecker delta.
 
 An important subset of  ${\mathbb C}[\lambda^{-1},\lambda]]$ is the field of real series ${\mathbb R}[\lambda^{-1},\lambda]].$ Its elements are of the form (\ref{uzup1}) but now every number $c_l$ is  assumed to be real.
 
 \subsection{Formal series of functions containing negative powers of $\lambda$}

The extension of the set of complex numbers to ${\mathbb C} [\lambda^{-1},\lambda]]$ implies a deep modification of the set of formal series $C^{\infty}[[\lambda]]({\cal M}).$ Namely we have to add to this collection of series also the formal Laurent series with respect to $\lambda,$ which principal part is finite
\be
\label{1..1}
\varphi[[\lambda]]:= \sum_{l=-u}^{\infty} \lambda^l \varphi_l\; , \;\;\; \forall \,l \;\; \varphi_l \in C^{\infty}({\cal M}).
\ee

    The generalisation of summation (\ref{a1}) for such series is straightforward.

  In this set denoted by $C^{\infty}[\lambda^{-1},\lambda]]({\cal M})$
the  multiplication  by a scalar belonging to ${\mathbb C} [\lambda^{-1},\lambda]]$ is of the following shape  
\[
\cdot : {\mathbb C} [\lambda^{-1},\lambda]] \times C^{\infty}[\lambda^{-1},\lambda]]({\cal M}) \rightarrow C^{\infty}[\lambda^{-1},\lambda]]({\cal
M})
\]
\be
\label{potrzeba} 
\sum_{l=-u}^{\infty} \lambda^l c_l \cdot \sum_{k=-s}^{\infty}\lambda^k \varphi_k= \sum_{l=0}^{\infty} \lambda^{l-u-s}
\sum_{k=0}^{l} c_{k-u} \varphi_{l-k-s}.
 \ee
  Since the parameter $\lambda$ is real, the complex conjugations are respectively 
  \[
   {\mathbb C}
[\lambda^{-1},\lambda]] \ni \overline{\sum_{l=-u}^{\infty} \lambda^l c_l}= \sum_{l=-u}^{\infty} \lambda^l \overline{c}_l\;\;\; {\rm and} \;\;\;
C^{\infty}[\lambda^{-1},\lambda]]({\cal M}) \ni \overline{\sum_{l=-s}^{\infty}\lambda^l \varphi_l}= \sum_{l=-s}^{\infty}\lambda^l
\overline{\varphi}_l. 
\]
 The set $(C^{\infty}[\lambda^{-1},\lambda]]({\cal M}),{\mathbb C} [\lambda^{-1},\lambda]],+,\cdot)$ of all formal series
$\varphi[[\lambda]]$ constitutes a linear space over the field ${\mathbb C} [\lambda^{-1},\lambda]].$

Let us propose  convergence of a sequence $\{ \varphi_n[[\lambda]] \}_{n=1}^{\infty}$ of formal series. By definition the sequence
$\{ (\sum_{l=-u}^{\infty}\lambda^l\varphi_l)_n \}_{n=1}^{\infty}$ tends to a series $\sum_{l=-u}^{\infty} \lambda^l \varphi_{l0},$ if
for every $l$ the sequence $\{ (\varphi_l)_n \}_{n=1}^{\infty}$ is convergent to the function $\varphi_{l0}$ in the sense of
convergence in the space $C^{\infty}({\cal M}).$

The formal series are differentiable. For every set of indices $ m_1, \ldots, m_{2r} \in {\cal N}, \; \dim {\cal M}=2r $ and every
formal series $\sum_{l=-u}^{\infty}\lambda^l \varphi_l$ we define its partial derivative as 
\be 
\label{2.1}
 \frac{
\partial^{m_1+m_2+\ldots + m_{2r}}}{\partial^{m_1}q^1 \ldots \partial^{m_{2r}}q^{2r} } \sum_{l=-u}^{\infty}\lambda^l \varphi_l :=
\sum_{l=-u}^{\infty}\lambda^l \frac{ \partial^{m_1+m_2+\ldots + m_{2r}}}{\partial^{m_1}q^1 \ldots \partial^{m_{2r}}q^{2r} } \varphi_l.
\ee 
There exists also a natural way of introducing an integration in the space
$C^{\infty}[\lambda^{-1},\lambda]]({\cal M}).$ Namely for every formal series $\sum_{l=-u}^{\infty}\lambda^l \varphi_l$ in which all functions $\varphi_l $ are summable
\be
 \label{2.2}
  \int_{\cal M}\left( \sum_{l=-u}^{\infty}\lambda^l \varphi_l \right) \,
\omega^r:= \sum_{l=-u}^{\infty}\lambda^l \int_{\cal M} \varphi_l \, \omega^r. 
\ee

\subsection{A trivial generalisation of multiplication of formal series}

The next step towards proposing a `really' deformed algebra of formal series is
some trivial generalisation of the pointwise multiplication of functions to 
formal series 
 \[
  \bullet:C^{\infty}[\lambda^{-1},\lambda]]({\cal M}) \times C^{\infty}[\lambda^{-1},\lambda]]({\cal M})
\rightarrow C^{\infty}[\lambda^{-1},\lambda]]({\cal M})
 \] 
given by 
\be 
\label{3}
 \sum_{l=-u}^{\infty}\lambda^l \varphi_l \bullet
\sum_{k=-s}^{\infty}\lambda^k \psi_k=\frac{1}{\lambda^{u+s}} \sum_{l=0}^{\infty}\lambda^l \sum_{k=0}^{l} \varphi_{k-u} \psi_{l-k-s}. 
\ee
The $\bullet$-- multiplication  is  of course Abelian and associative.

  The set of series
$C^{\infty}[\lambda^{-1},\lambda]]({\cal M})$ with the product `$\bullet$' is a commutative ring. It is also an algebra over the field ${\mathbb
C} [\lambda^{-1},\lambda]].$ The complex conjugation is an involution in this algebra.
Moreover the $\bullet$--product is continuous in its arguments.
If
$
\{ \varphi_n[[\lambda]] \}_{n=1}^{\infty} \rightarrow \sum_{l=-u}^{\infty} \lambda^l \varphi_{l0} \;\; {\rm and} \;\;
\{ \psi_n[[\lambda]] \}_{n=1}^{\infty} \rightarrow \sum_{l=-s}^{\infty} \lambda^l \psi_{l0} \;\; {\rm then} 
$
\[
\lim_{j,k \rightarrow \infty} \Big(
\{ \varphi_j[[\lambda]] \}_{j=1}^{\infty} \bullet \{ \psi_k[[\lambda]] \}_{k=1}^{\infty} \Big)=
\sum_{l=-u}^{\infty} \lambda^l \varphi_{l0} \bullet \sum_{l=-s}^{\infty} \lambda^l \psi_{l0}.
\]
 The $\bullet$-- multiplication
satisfies the Leibniz product rule
 \[
  \forall \;1 \leq s \leq 2r \;\; \forall \; \varphi[[\lambda]], \psi[[\lambda]] \in
C^{\infty}[\lambda^{-1},\lambda]]({\cal M}) 
\] 
\be 
\label{3.1} \;\; \frac{\partial}{\partial q^s} \Big(\varphi[[\lambda]] \bullet \psi[[\lambda]]
\Big) = \frac{\partial \, \varphi[[\lambda]]}{\partial q^s} \bullet \psi[[\lambda]] + \varphi[[\lambda]] \bullet \frac{\partial \,
\psi[[\lambda]]}{\partial q^s}.
 \ee 
The integration (\ref{2.2}) becomes a trace in the algebra
$(C^{\infty}[\lambda^{-1},\lambda]]({\cal M}), \bullet)$ because for every two formal series of functions with compact supports $\varphi[[\lambda]],
\psi[[\lambda]]$ \[ \int_{\cal M} \varphi[[\lambda]] \bullet \psi[[\lambda]] \, \omega^r = \int_{\cal M} \psi[[\lambda]] \bullet
\varphi[[\lambda]] \, \omega^r. \]

\subsection{A $*$ -- product of formal series }
One of the fundamental differences between the sets of classical and of quantum observables is a way in which we multiply their elements. Both of these
multiplications are associative but only the quantum one is non-Abelian. The ring of series $C^{\infty}[\lambda^{-1},\lambda]]({\cal M})$ with the
product `$\bullet$' contains the deformation parameter $\lambda$ but the $\bullet$--product in it is Abelian. Thus the set
$C^{\infty}[\lambda^{-1},\lambda]]({\cal M})$ is  the `ordinary' space $C^{\infty}({\cal M})$ with the deformation parameter
$\lambda$  put by hand. A real step towards modelling the space of quantum observables requires a nontrivial, which
especially means non-Abelian, deformation of the $\bullet$--multiplication.
  
Thus we introduce a $*$--product being the mapping 
\[ 
*: C^{\infty}[\lambda^{-1},\lambda]]({\cal M}) \times C^{\infty}[\lambda^{-1},\lambda]]({\cal M})
\rightarrow C^{\infty}[\lambda^{-1},\lambda]]({\cal M}) 
\] 
which for functions $\varphi, \psi \in C^{\infty}({\cal M}) $ is of the form 
\be
\label{3.101} 
\varphi * \psi := \sum_{k=0}^{\infty}\lambda^k B_k(\varphi,\psi),\;\; \forall \; k \;\; B_k(\varphi,\psi) \in
C^{\infty}({\cal M}) .
 \ee
  Several conditions are imposed on the $*$--product. They arise from  demand, that this multiplication
should have properties analogous to the product of linear  operators acting in a Hilbert space.
\begin{enumerate}
 \item 
 The operators
$B_k(\cdot,\cdot), \;\; k=0,1,2, \ldots$ are ${\mathbb C}[\lambda^{-1},\lambda]]$--bilinear.
 \item 
 The operators $B_k(\cdot,\cdot), \;\;
k=0,1,2, \ldots$ are local i.e. 
\[ 
\forall \; \varphi, \psi \in C^{\infty}({\cal M}) \;\;\; {\rm supp} \; B_k(\varphi,\psi) \subset
\big( {\rm supp} \;\varphi \; \cap {\rm supp} \;\psi \big).
 \] 
 \item 
 \label{it1}
 The $*$--product is associative. Thus for every $k
\geq 0$ and every functions $\varphi, \psi, \phi \in C^{\infty}({\cal M})$ 
\[ 
\sum_{l=0}^k B_l(\varphi, B_{k-l}(\psi, \phi))=
\sum_{l=0}^k B_l(B_{k-l}(\varphi,\psi),\phi). 
\]
 \item
  For every $\varphi, \psi \in C^{\infty}({\cal M}) $ there is
   \[ 
   B_0(\varphi,
\psi)= \varphi \cdot \psi. 
\] 
This requirement means that the $*$--product is indeed a deformation of both: the pointwise multiplication of
functions and the $\bullet$--product of formal series. 
\item 
For every $k \geq 1$ and every function $\varphi \in C^{\infty}({\cal M})
$ there is 
\[ 
B_k({\bf 1},\varphi) = B_k(\varphi,{\bf 1})=0.
 \] 
 Therefore the constant function equal to ${\bf 1}$ is the identity
element with respect to the $*$--product. 
\item 
For all $ \varphi, \psi \in C^{\infty}({\cal M})$ \[ B_1(\varphi, \psi)- B_1(\psi,
\varphi)= i \{ \varphi, \psi \}_P, \] where $ \{ \cdot, \cdot \}_P$ symbolises the Poisson bracket. It ensures that in the first
approximation with respect to the deformation parameter $\lambda$ the commutator 
\[
 (\varphi * \psi- \psi* \varphi ) \sim i \lambda \{
\varphi, \psi \}_P.
 \]
  \item 
  The complex conjugation is an involution in the $*$--algebra i.e. for every $\varphi, \psi \in
C^{\infty}({\cal M}) $ 
\[ 
\overline{B_k(\varphi, \psi)}= B_k(\overline{\psi}, \overline{\varphi}). 
\] 
Thus we say that the
$*$--product is Hermitian. 
\end{enumerate} 

There are two more supplementary assumptions about the $*$--product, which are widely used.
\begin{enumerate}
 \setcounter{enumi}{7} 
 \item 
 The operators $B_k(\cdot,\cdot), \;\; k=0,1,2, \ldots$ are bidifferential. This
condition assures a convenient realisation of the previous requirements. 
\item
 Every bidifferential operator $B_k(\cdot,\cdot), \;\;
k=0,1,2, \ldots$ is at most of the order $k.$ In this case we call the $*$--product natural. 
\end{enumerate}

 From these assumptions we
see that for arbitrary formal series $\sum_{l=-u}^{\infty}\lambda^l \varphi_l, \sum_{k=-s}^{\infty}\lambda^k \psi_k \in
C^{\infty}[\lambda^{-1},\lambda]]({\cal M})$ their $*$--product 
\be 
\label{3.2}
 \sum_{l=-u}^{\infty}\lambda^l \varphi_l *
\sum_{k=-s}^{\infty}\lambda^k \psi_k =\frac{1}{\lambda^{u+s}} \sum_{l=-u}^{\infty}
\sum_{k=-s}^{\infty}\sum_{m=0}^{\infty}\lambda^m B_m(\varphi_l, \psi_k).
\ee 
The set of series
$C^{\infty}[\lambda^{-1},\lambda]]({\cal M})$ with the $*$--product is a ring, which in general is non--Abelian. Moreover, it is an algebra over
the field ${\mathbb C} [\lambda^{-1},\lambda]].$ The $*$--product is continuous in its arguments if all of the operators $B_k(\cdot, \cdot)$ are
continuous. Thus every natural $*$--multiplication is continuous. On the contrary to the $\bullet$--product, the $*$--product may not
obey the Leibniz product rule.

An interesting question is existence and uniqueness of  trace in the algebra $(C^{\infty}[\lambda^{-1},\lambda]]({\cal M}),*).$ 
This functional plays the fundamental role in calculating   mean values of observables. 
As it was proved
\cite{7, nest, gutt},  for every differential $*$-- product there exists a unique, up to  multiplication (\ref{potrzeba}) by a
constant from ${\mathbb C}[\lambda^{-1},\lambda]],$ trace determined by a density $t[[\lambda]] \in C^{\infty}[\lambda^{-1},\lambda]]({\cal M})$ such that for every two
formal series $\varphi[[\lambda]], \psi[[\lambda]]$ with compact supports 
\be 
\label{3.332}
 \int_{\cal M} \Big(  \varphi[[\lambda]] *
\psi[[\lambda]] \Big)\bullet t[[\lambda]] \, \omega^r = \int_{\cal M} \Big(  \psi[[\lambda]] * \varphi[[\lambda]] \Big) \bullet t[[\lambda]] \, \omega^r. 
\ee
 Therefore as the trace we choose
 \be 
 \label{3.333}
  {\rm Tr \,
\varphi[[\lambda]]}:= \frac{1}{\lambda^r} \int_{\cal M} \varphi[[\lambda]] \bullet t[[\lambda]] \, \omega^r. 
\ee
 The coefficient $
\frac{1}{\lambda^r}$ follows from the definition of  trace in  algebra $(C^{\infty}[\lambda^{-1},\lambda]]({\mathbb R}^{2r}),*_{ M})$ with the Moyal $*$--product. We will consider this algebra later.

The observation that the trace density $t[[\lambda]]$ is defined up to a constant implies, that without any loss of generality one can assume  $t[[\lambda]]= \sum_{k=0}^{\infty} \lambda^k t_k.$

Another problem is a covariant way of defining the $*$--multiplication. We are not going to discuss this question.

Practical realizations of the $*$--product are well known. On the manifold ${\mathbb R}^{2r}$ the mentioned bidifferential $*$--product called
 Moyal product was introduced by H. J. Groenewold \cite{GW46} and J. E. Moyal \cite{MO49}. An extension of this product on the case
of an arbitrary symplectic manifold was proposed by B. Fedosov \cite{6, 7} and  construction of the $*$--product on a Poisson
manifold was found by M. Kontsevich \cite{kon}.

Theory of deformation is a dynamically developing discipline of mathematics. However, it remains an open question, which of these
known $*$--products are really useful in physical applications. 


\section{Linear continuous functionals as states}
\setcounter{equation}{0}

A set of observables itself is not sufficient to represent physical reality. The second basic component of the mathematical model of reality are
states \cite{zy}. This section contains a detailed analysis  of  the spaces of states for three algebras: the classical
one $(C^{\infty}({\cal M}), \cdot),$ the trivially deformed one $(C^{\infty}[\lambda^{-1},\lambda]]({\cal M}), \bullet) $ and the quantum one
$(C^{\infty}[\lambda^{-1},\lambda]]({\cal M}), *). $ 

\subsection{Classical states}

  In classical physics we identify the states with these linear continuous functionals over the space of
functions $C^{\infty}({\cal M}),$ which are real 
\setcounter{orange}{1} \renewcommand{\theequation}{\arabic{section}.\arabic{equation}\theorange} 
\be
 \label{3.21}
  \forall \, \varphi \in C^{\infty}({\cal M})\;\; {\rm such\;\; that }\;\;
\overline{\varphi}= \varphi\;\;\;{\rm there \;\; is}\;\; \big< T, \varphi \big> = \overline{\big< T, \varphi \big>}, 
\ee
 nonnegative
\addtocounter{orange}{1} \addtocounter{equation}{-1} 
\be 
\label{3.3}
 \forall \, \varphi \in C^{\infty}({\cal M})\;\; \big< T,
\overline{\varphi} \varphi\big> \geq 0 
\ee 
and satisfy the normalisation condition
 \addtocounter{orange}{1}
\addtocounter{equation}{-1} 
\be 
\label{3.4} 
\big<T, {\bf 1} \big>=1.
 \ee 
 \renewcommand{\theequation}{\arabic{section}.\arabic{equation}} 
 Thus we see that the classical states may be identified with some subset of generalised functions
with compact supports. The space of  functionals 
with compact supports
will be denoted by ${\cal E}'({\cal M}).$ 

Notice that the condition (\ref{3.3}) is not equivalent to the requirement that $ \big< T, \psi \big> \geq 0 $ for every nonnegative function $\psi \in C^{\infty}({\cal M})$ because not every nonnegative function  is the square of the absolute value of a smooth function ( e.g.  $\psi= x^2+p^2$ cannot be written as $\psi=|\phi|^2,$ where $\phi \in C^{\infty}({\cal M})$). We propose (\ref{3.3}) in order to have a formula analogous to the one appearing for  quantum algebra
$(C^{\infty}[\lambda^{-1},\lambda]]({\cal M}), *).$ It seems that our choice does not imply any ambiguities. 

In the case when the functional $T$ is an integrable function, we identify the functional action with an integral according to the
rule
 \be
 \label{03}
 \big< T, \varphi \big> = \int_{\cal M}T\, \varphi\, \omega^r. 
 \ee
 If the symplectic manifold $({\cal M}, \omega)$ is $({\mathbb R}^2, dq \wedge dp),$ an example of state is
 the Dirac delta $\delta(p-{\tt p}_0) \otimes \delta(q-{\tt q}_0).$ 
 Also every nonnegative normalised summable function with compact support defined on the manifold ${\cal M}$ represents a state.
  
  A sequence of linear functionals $\{T_n\}_{n=1}^{\infty}$ belonging to ${\cal E}'({\cal M})$ is
convergent to a functional $T_0$ if all ${\rm supp}\; T_n$ are contained in a common compact set and
\be
\label{003}
 \forall \; \varphi \in C^{\infty}({\cal M}) \; \lim_{n \rightarrow \infty} \big<T_n, \varphi
\big>= \big<T_0, \varphi \big>. 
\ee
Every generalised function belonging to ${\cal E}'({\cal M})$ is a limit in the sense of (\ref{003}) of a sequence of smooth functions with compact supports.

Possible results of  measurement of  an 
observable $\varphi$ are numbers from the range of the function $\varphi.$ Although it is not considered very often in classical physics, one can introduce an eigenvalue equation for a quantity $\varphi$ 
\be
\label{0003}
\forall \; \psi \in C^{\infty}({\cal M}) \;\;\;\;\; \big<\varphi \cdot T_a, \psi \big>
 =  \big<a T_a,\psi \big>
\ee
with
$
\big<\varphi \cdot T_a, \psi \big>:= \big<T_a, \varphi \cdot \psi \big>.
$
By $a$ we denote an eigenvalue of $\varphi,$ the product $\varphi \cdot T_a$ symbolises the multiplication of the generalised function $T_a$ by the smooth function $\varphi.$ Every   generalised function $T_a$ satisfying (\ref{0003}) and the definition of  classical state is an eigenstate of the observable $\varphi$ related to the eigenvalue $a.$ As in quantum mechanics, eigenvalues of the quantity $\varphi$ are possible results of its measurement and eigenstates $T_a$ are states, at which we certainly obtain the result $a.$  

 Notice that the definition of state introduced in this subsection is very restrictive. Thus e. g. the Gaussian distribution of probability on the space ${\mathbb R}^{2n}$ does not belong to it. However, since  ${\cal M}$ is an arbitrary symplectic manifold, it would be difficult to propose  any classes of test functions different from the set $C^{\infty}({\cal M}).$
\subsection{States of the trivially deformed observables}

Let us consider  construction of a space of states over the commutative ring $( C^{\infty}[\lambda^{-1},\lambda]]({\cal M}), \bullet).$ Although this example has
no special physical meaning, it illustrates very well some difficulties, which appear when we deal with formal series. In particular the notion of nonnegativity and an eigenvalue
problem are worth  analysis.
 
The extension of the space $C^{\infty}({\cal M})$ of smooth functions to the space of formal series $C^{\infty}[\lambda^{-1},\lambda]]({\cal M})$
implies some modifications of the space ${\cal E}'({\cal M})$ of linear and continuous functionals over $C^{\infty}({\cal M}).$ First
of all it is necessary to extend the action of any functional $T \in {\cal E}'({\cal M})$ on every formal series (\ref{1..1}). By
definition 
\be
 \label{3.5}
  \Big<T, \sum_{l=-u}^{\infty}\lambda^l \varphi_l \Big>:= \sum_{l=-u}^{\infty}\lambda^l \Big<T, \varphi_l
\Big>\; \in {\mathbb C} [\lambda^{-1},\lambda]]. 
\ee 
The  ${\mathbb C} [\lambda^{-1},\lambda]]$--linearity of  functional $T \in {\cal E}' ({\cal M})$ over the space
$C^{\infty}[\lambda^{-1},\lambda]]({\cal M})$ means that
 \[ 
 \forall \; c_1[[\lambda]], c_2[[\lambda]] \in {\mathbb C} [\lambda^{-1},\lambda]], \;\;\; \forall \; \varphi_1[[\lambda]],
\varphi_2[[\lambda]] \in C^{\infty}[\lambda^{-1},\lambda]]({\cal M})\;\; 
\]
\be
\label{3.55}
\Big< T, c_1[[\lambda]] \cdot \varphi_1[[\lambda]] + c_2[[\lambda]] \cdot \varphi_2[[\lambda]] \Big> = c_1[[\lambda]] \Big< T, \varphi_1[[\lambda]]
\Big> + c_2[[\lambda]] \Big< T, \varphi_2[[\lambda]] \Big>.
\ee
Every linear functional over $C^{\infty}({\cal M})$ extended to a functional over $C^{\infty}[\lambda^{-1},\lambda]]({\cal M})$ according to the rule (\ref{3.5}) obeys (\ref{3.55}).

Let us prove  continuity of the functional $T$ defined by (\ref{3.5}) with respect to the topology of $ C^{\infty}[\lambda^{-1},\lambda]]({\cal M}).$ Assume that $\{
(\sum_{l=-u}^{\infty}\lambda^l\varphi_l)_n \}_{n=1}^{\infty} \longrightarrow \sum_{l=-u}^{\infty} \lambda^l \varphi_{l0}.$ Then 
\[ 
\Big< T, (\sum_{l=-u}^{\infty}\lambda^l\varphi_l)_n \Big>=
\sum_{l=-u}^{\infty} \lambda^l \Big< T, (\varphi_{l})_n \Big> \longrightarrow \sum_{l=-u}^{\infty} \lambda^l \Big< T, (\varphi_{l})_0
\Big> = \Big< T, \sum_{l=-u}^{\infty}\lambda^l\varphi_{l0} \Big>.
 \] 
 Thus we see that linear continuous functionals over $C^{\infty}({\cal M})$ can be extended in a natural way to linear and continuous functionals over the space  $C^{\infty}[\lambda^{-1},\lambda]]({\cal M}).$
 
 Next we discuss the extra requirements imposed on functionals representing states. 
 The condition of reality (\ref{3.21}) of a generalised functional has a straightforward
extension for formal series. We say that a linear functional $T \in {\cal E}'({\cal M})$ is real over $C^{\infty}[\lambda^{-1},\lambda]]({\cal M})$, if for every formal series 
$\sum_{l=-u}^{\infty}\lambda^l \varphi_l $ the implication holds
 \be
 \label{4.9}
\sum_{l=-u}^{\infty}\lambda^l \overline{ \varphi_l }=\sum_{l=-u}^{\infty}\lambda^l \varphi_l
\Longrightarrow
 \Big<T,
\sum_{l=-u}^{\infty}\lambda^l \varphi_l \Big> \in {\mathbb R}[\lambda^{-1}, \lambda]]. 
\ee
Thus we see that a linear functional  $T \in {\cal E}'({\cal M})$ is real over the space  $C^{\infty}[\lambda^{-1},\lambda]]({\cal M})$ if and only if it is real over the space $C^{\infty}({\cal M}).$

Reformulation of  definition of nonnegativity
(\ref{3.3}) for functionals over the formal series belonging to $C^{\infty}[\lambda^{-1},\lambda]]({\cal M})$ is required. The notion of nonnegativity is in conflict with the idea of formal series. On one hand we deal with specific real numbers. On the other hand we avoid the question about summability. In our opinion there is no satisfactory resolution of this dilemma.
What we propose is a compromise between intuition and mathematical elegance.

Since nonnegativity refers to the value of functional action we need to consider all possible numerical values of $\lambda.$ 
  \begin{de}
\label{positivity2}
A generalised function $ T \in {\cal E}'({\cal M})$ is nonnegatively defined over the ring $ (C^{\infty}[\lambda^{-1},\lambda]]({\cal M}), \bullet)$,
if for every admissible value of the parameter $\lambda$  and every finite sum
$\sum_{l=-u}^{s}\lambda^l \varphi_l$ 
\be
 \label{5}
  \Big< T, \sum_{k_1=-u}^{s}\lambda^{k_1} \overline{\varphi}_{k_1} \bullet
\sum_{k_2=-u}^{s}\lambda^{k_2} \varphi_{k_2} \Big> \geq 0. 
\ee 
\end{de}
We restrict to the finite sum because of problems with
convergence of formal expressions. 

Our definition of nonnegativity is  stronger than the one proposed by S. Waldmann \cite{wald}. We are motivated by the fact that a concrete value of the deformation parameter depends on a choice of units and can be arbitrarily large. 

For any nonnegative value of $\lambda$ and any $s \in {\cal N}$ the linear combination $\sum_{l=-u}^{s}\lambda^l \varphi_l $ is a function belonging to $
C^{\infty}({\cal M}).$ 
Vice versa, every function $\varphi \in C^{\infty}({\cal M}) $ for an arbitrary  fixed value of $\lambda$ and  arbitrary fixed numbers $u,\, s$ may be written as a sum $\sum_{l=-u}^{s}\lambda^l \varphi_l $ and this representation is not unique.

Therefore  the condition (\ref{5}) leads to the following observation.
\begin{co} 
\label{corollary1}
A generalised function $T \in {\cal
E}'({\cal M}) $ is nonegatively defined over the ring $ (C^{\infty}[\lambda^{-1},\lambda]]({\cal M}), \bullet)$ if and only if it is nonnegatively
defined over the ring $ (C^{\infty}({\cal M}), \cdot).$ 
\end{co} 
Definition \ref{positivity2} refers to numerical values of the deformation parameter which is against the spirit of formal series calculus. Hence it would be welcome to reformulate  the meaning of nonegativity in a form not mentioning  possible values of $\lambda$ and  operating on arbitrary formal series. At first glance a sufficient condition for a generalised function $T  \in {\cal E}'({\cal M})$ to be nonnegative could state that
  for every natural number $l$ fulfilling the inequality $ - 2u \leq l$
\be
\label{5.511}
\lambda^l \Big< T,  \sum_{k_1 + k_2=l} \overline{\varphi}_{k_1} \varphi_{k_2}\Big> \geq 0.
\ee
Recall that $-u \leq k_1, k_2.$

Unfortunately, one  concludes that
\begin{co}
If a generalised function $T  \in {\cal E}'({\cal M}) $  for every sum $\sum_{l=-u}^{s}\lambda^l \varphi_l$ satisfies the conditions (\ref{5}) and (\ref{5.511}) then the functional action 
\[
\Big<T,\sum_{k_1=-u}^{s}\lambda^{k_1}\overline{ \varphi}_{k_1} \bullet \sum_{k_2=-u}^{s}\lambda^{k_2} \varphi_{k_2} \Big> =0
\]
 so $T$ cannot be a state.
\end{co}
Indeed, let us consider two sums: $\varphi_0$ and $\varphi_0 -  \lambda \varphi_0, $  where $\varphi_0 \in C^{\infty}({\cal M}).$ Applying (\ref{5}) to function $\varphi_0$ we obtain 
\[
\Big< T,  \overline{\varphi}_{0} \varphi_{0} \Big> \geq 0. 
\]
 On the other hand the requirement (\ref{5.511}) for $l=1$ leads  to the conclusion
\[
\lambda  \Big< T, \overline{\varphi}_{0} \varphi_{0} \Big> \leq 0. 
\]
Since the function $\varphi_{0}$ is arbitrary and $\lambda $ is positive, we see that for every $\varphi_0 \in C^{\infty}({\cal M})$ the equality holds
\[
\Big< T,  \overline{\varphi}_{0} \varphi_{0} \Big> = 0. 
\]
Therefore from Corollary \ref{corollary1} we obtain that  $\Big<T,\sum_{k_1=-u}^{s}\lambda^{k_1}\overline{ \varphi}_{k_1} \bullet \sum_{k_2=-u}^{s}\lambda^{k_2} \varphi_{k_2} \Big>=0 $ for every sum $\sum_{l=-u}^{s}\lambda^{l}\overline{ \varphi}_{l}.$ It especially means that 
\[
\Big< T,  \overline{\bf 1} \cdot {\bf 1} \Big>= \Big< T,  {\bf 1} \Big>=0
\]
so $T$ cannot be normalised and thus it is not a state.
 \rule{2mm}{2mm}
 
 An equivalent point of view at nonnegativity would be based on the following definition.
 \begin{de}
 \label{pos4}
 A formal series $\sum_{l=-u}^{\infty} \lambda^l c_l \; \in {\mathbb R}[\lambda^{-1}, \lambda]]$ is nonnegative if 
 \[
 \forall\; \lambda>0 \;\exists \, k \in {\cal N}\;  \forall \; m >k \; \sum_{l=-u}^{m} \lambda^l c_l  \geq 0.
 \]
 \end{de}
 With its use one can say that
   \begin{de}
\label{positivity4}
A generalised function $ T \in {\cal E}'({\cal M})$ is nonegatively defined over the ring $ (C^{\infty}[\lambda^{-1},\lambda]]({\cal M}), \bullet)$,
if for every admissible value of  parameter $\lambda$  and every formal  series
$\sum_{l=-u}^{\infty}\lambda^l \varphi_l$ 
the formal series of real numbers
\[
  \Big< T, \sum_{k_1=-u}^{\infty}\lambda^{k_1} \overline{\varphi}_{k_1} \bullet
\sum_{k_2=-u}^{\infty}\lambda^{k_2} \varphi_{k_2} \Big> 
\]
is nonnegative.
\end{de}

Finally, the normalisation condition (\ref{3.4}) does not require any special adaptation to formal series. 
 
Thus we get the subset of the space of linear continuous functionals over the algebra $( C^{\infty}[\lambda^{-1},\lambda]]({\cal M}), \bullet),$
which elements represent states. As expected, this space may be identified with the set of classical states. 
 
However, it seems to be natural that the deformation parameter $\lambda$  impacts also the space of generalised functions over $
C^{\infty}[\lambda^{-1},\lambda]]({\cal M})$. There are several possible extensions of the set ${\cal E}'({\cal M}) $ with respect to $\lambda.$ In the first step 
we propose a new space of functionals $ {\cal E}' [\lambda^{-1},\lambda]] ({\cal M})$
     isomorphic to the Cartesian product 
 \[    
     {\cal E}' [\lambda^{-1},\lambda]] ({\cal M})= 
     {\mbox {\Large {$\times$}}}_{l=0}^{\infty} {\cal
E}'_l({\cal M}), \;\; \forall \; l \;\; {\cal E}'_l({\cal M}):= {\cal E}'({\cal M}). 
\]
 One can verify that the elements of this new
space are ${\mathbb C} [\lambda^{-1},\lambda]]$--linear and continuous functionals over the space $C^{\infty}[\lambda^{-1},\lambda]]({\cal M})$ with the
action defined in the following way 
\[
\forall \; \sum_{l=-s}^{\infty}\lambda^l T_l \in {\cal E}' [\lambda^{-1},\lambda]] ({\cal M})\;\; {\rm and}\;\;
\forall \; \sum_{k=-u}^{\infty} \lambda^k \varphi_k \in C^{\infty}[\lambda^{-1},\lambda]]({\cal M})
\]
\be 
\label{4} 
\Big< \sum_{l=-s}^{\infty}\lambda^l T_l,\sum_{k=-u}^{\infty}\lambda^k \varphi_k\Big> := \frac{1}{\lambda^{s+u}} \sum_{z=0}^{\infty} \lambda^z
\sum_{l=0}^{z}\Big< T_{l-s}, \varphi_{z-l-u} \Big> \in {\mathbb C} [\lambda^{-1},\lambda]]. 
\ee
This formula is consistent with (\ref{3.5}).
 ${\mathbb C} [\lambda^{-1},\lambda]]$--linearity of the proposed action is obvious. 
 
 We add one extra condition imposed on the set of supports of distributions $T_l.$ It is required  that all of those supports are contained in a common compact set. Explanation of this condition is the following. For some numerical values of $\lambda$ the series $\sum_{l=-s}^{\infty}\lambda^l T_l$ can be convergent in the sense of convergence of generalised functions. The proposed constraint ensures that then the sum $\sum_{l=-s}^{\infty}\lambda^l T_l$ belongs to ${\cal E}'({\cal M}). $

Let us prove the continuity of an arbitrary functional $\sum_{l=-s}^{\infty}\lambda^l T_l \in {\cal E}' [\lambda^{-1},\lambda]] ({\cal M})$ with
respect to the topology of $ C^{\infty}[\lambda^{-1},\lambda]]({\cal M}).$ As before, assume that $\{ (\sum_{k=-u}^{\infty}\lambda^k\varphi_k)_n
\}_{n=1}^{\infty} \longrightarrow \sum_{k=-u}^{\infty} \lambda^k \varphi_{k0}.$ 
Then for every formal series
$\sum_{l=-s}^{\infty}\lambda^l T_l$
\[ 
\Big< \sum_{l=-s}^{\infty}\lambda^l T_l, (\sum_{k=-u}^{\infty}\lambda^k\varphi_k)_n \Big>=
\frac{1}{\lambda^{s+u}}
\sum_{z=0}^{\infty} \lambda^z \sum_{l=0}^{z}\Big< T_{l-s}, (\varphi_{z-l-u})_n \Big> \longrightarrow
\] 
\[
 \longrightarrow
 \frac{1}{\lambda^{s+u}}
\sum_{z=0}^{\infty} \lambda^z \sum_{l=0}^{z}\Big< T_{l-s}, (\varphi_{z-l-u})_0 \Big> = \Big< \sum_{l=-s}^{\infty}\lambda^l T_l,
\sum_{k=-u}^{\infty}\lambda^k\varphi_{k0} \Big>, 
\]
 because for every $l$ the functional $T_l$ is a generalised function over $
C^{\infty}({\cal M}).$ 

Reality of a formal series $\sum_{l=-s}^{\infty} \lambda^l T_l $  is understood as requirement that for every 
$\sum_{k=-u}^{\infty}\lambda^k\varphi_{k} \in  C^{\infty}[\lambda^{-1},\lambda]]({\cal M})$ there is
\be
\label{5.555}
\sum_{k=-u}^{\infty}\lambda^k\overline{\varphi_{k}}=\sum_{k=-u}^{\infty}\lambda^k\varphi_{k} \Longrightarrow
\Big<\sum_{l=-s}^{\infty} \lambda^l T_l,  \sum_{k=-u}^{\infty}\lambda^k\varphi_{k}\Big>  \in {\mathbb R }[\lambda^{-1},\lambda]].
\ee

 In the next step we extend the definition (\ref{5}) of nonnegativity to the formal series of generalised
functions. 
\begin{de} 
A formal series $\sum_{l=-s}^{\infty} \lambda^l T_l \in {\cal E}'[\lambda^{-1},\lambda]]({\cal M})$ is nonegatively defined if for every series $ \sum_{m=-u}^{\infty} \lambda^{m} \varphi_{m} \in C^{\infty}[\lambda^{-1},\lambda]]({\cal M})$
the formal series of real numbers
\be 
\label{5.5}
 \Big< \sum_{l=-s}^{\infty} \lambda^l T_l, \sum_{m_1=-u}^{\infty}
\lambda^{m_1} \overline{\varphi}_{m_1} \bullet \sum_{m_2=-u}^{\infty} \lambda^{m_2} \varphi_{m_2}\Big> 
 \ee 
 is nonnegative in the sense of Def. \ref{pos4}.
\end{de}
 Keeping in mind Corollary \ref{corollary1} we
conclude that
 \begin{co} 
 \label{cowazny}
 A formal series $\sum_{l=-s}^{\infty} \lambda^l T_l \in {\cal E}'[\lambda^{-1},\lambda]]({\cal M})$ is nonnegatively defined
if for every positive value of $\lambda$ and 
every 
test function $ \varphi \in C^{\infty}({\cal M})$ 
there exists a number $k \in {\cal N} $ such that for every natural number $u>k$  
\[ 
\Big< \sum_{l=-s}^{u} \lambda^l T_l, \overline{\varphi} \varphi \Big> \geq 0. 
\]
\end{co}
The problem of nonnegativity of functionals over the trivially deformed algebra is interesting in itself but since our scope are  quantum states, we are not going to continue with it.

The condition of normalisation for a series $\sum_{l=-s}^{\infty} \lambda^l T_l $ leads to the following equation on a formal series of complex numbers  $A[[\lambda]]$
\be
\label{w1}
A[[\lambda]]\cdot
\big< \sum_{l=-s}^{\infty} \lambda^l T_l, {\bf 1} \big>=1.
\ee
Equation (\ref{w1}) can be solved with respect to $A[[\lambda]]$ by recurrence. Notice that the series $A[[\lambda]]$ is an inverse element to $\big< \sum_{l=-s}^{\infty} \lambda^l T_l, {\bf 1} \big> $ in the field ${\mathbb C} [\lambda^{-1},\lambda]].$

 For every normalised series $\sum_{l=-u}^{\infty} \lambda^l {\cal T}_l:= A[[\lambda]] \cdot \sum_{l=-s}^{\infty} \lambda^l T_l $ there is
\be
\label{w11}
\big< {\cal T}_0, {\bf 1} \big>=1\;\;, \;\; \forall\; k \neq 0\;\; \big< {\cal T}_k, {\bf 1} \big>=0.
\ee

To deal with an eigenvalue problem we introduce a $\bullet$ --product of a functional $\sum_{l=-s}^{\infty} \lambda^l T_l \in {\cal E}'[\lambda^{-1},\lambda]]({\cal M})$ and  a formal
series $ \sum_{m=-r}^{\infty} \lambda^{m} \varphi_{m} \in C^{\infty}[\lambda^{-1},\lambda]]({\cal M}).$ By definition this multiplication is calculated as
\[
\bullet: C^{\infty}[\lambda^{-1},\lambda]]({\cal M}) \times {\cal E}'[\lambda^{-1},\lambda]]({\cal M})
\longrightarrow {\cal E}'[\lambda^{-1},\lambda]]({\cal M})
\]
\be
\label{w2}
 \forall \;\; \psi \in C^{\infty}({\cal M})\;\; \Big<\sum_{m=-u}^{\infty} \lambda^{m} \varphi_{m} \bullet
\sum_{l=-s}^{\infty} \lambda^l T_l, \psi \Big>:= \Big< \sum_{l=-s}^{\infty} \lambda^l T_l, \sum_{m=-u}^{\infty} \lambda^{m} \varphi_{m}
\bullet \psi \Big> 
\ee
and follows directly from the idea of multiplication of a generalised function by a smooth function. It can be reduced to the well known multiplication of the generalised function by a smooth function
 \be
 \label{w3}
 \sum_{m=-u}^{\infty} \lambda^{m} \varphi_{m} \bullet
\sum_{l=-s}^{\infty} \lambda^l T_l = \frac{1}{\lambda^{u+s}} \sum_{m=0}^{\infty} \lambda^{m} \sum_{l=0}^{m} \varphi_{l-u} \cdot T_{m-l-s}.
 \ee
 \begin{de}
\label{def3.2} 
The eigenvalue equation in the algebra $ (C^{\infty}[\lambda^{-1},\lambda]]({\cal M}), \bullet)$ for a formal series $ \sum_{m=-u}^{\infty}
\lambda^{m} \varphi_{m} \in C^{\infty}[\lambda^{-1},\lambda]]({\cal M})$ is of the form
 \be
  \label{3.6} 
  \forall \;\; \psi \in C^{\infty}({\cal M})\;\;
\Big<\sum_{m=-u}^{\infty} \lambda^{m} \varphi_{m} \bullet \sum_{l=-s}^{\infty} \lambda^l T_{a\;l}, \psi \Big>:= \Big< a[[\lambda]] \cdot
\sum_{l=-s}^{\infty} \lambda^l T_{a\;l}, \psi \Big>, 
\ee 
where $a[[\lambda]] \in {\mathbb C}[\lambda^{-1},\lambda]]$ is an  eigenvalue of the formal
series $ \sum_{m=-u}^{\infty} \lambda^{m} \varphi_{m} \in C^{\infty}[\lambda^{-1},\lambda]]({\cal M})$ and  the state $\sum_{l=-s}^{\infty} \lambda^l
T_{a\;l}$ its respective eigenstate. 
\end{de}

  Notice that for a fixed value of $a[[\lambda]]$ every ${\mathbb C}[\lambda^{-1},\lambda]]$--linear combination of  solutions   is also a solution of (\ref{3.6}). 
  
  A necessary condition for $a[[\lambda]]$ to be an eigenvalue of formal  series 
  $ \sum_{m=-u}^{\infty} \lambda^{m} \varphi_{m} $ is that the element 
  \[
  \sum_{m=-u}^{\infty} \lambda^{m} \varphi_{m} - a[[\lambda]] \cdot {\bf 1}
  \]
   is irreversible in the algebra  $ (C^{\infty}[\lambda^{-1},\lambda]]({\cal M}), \bullet).$
  
 The condition (\ref{3.6}) is equivalent to the system of equations: 
\bea
\label{3.61}
 (\varphi_{-u} -
a_{-u})T_{-s} &= & 0, \nonumber \\ 
(\varphi_{-u} - a_{-u})T_{1-s}+ (\varphi_{1-u} - a_{1-u})T_{-s} &= & 0,  \\
 \vdots & \vdots & \vdots \nonumber
 \eea 
This system is extremely restrictive, because e.g. if ${\rm supp}\, \varphi_{-u} = {\cal M},$ and the function $\varphi_{-u}$ has no
plateau, then the support of the generalised function $T_{-s}$ will be of measure zero. In a similar way one can deduce
that the analogous obstacle appears for higher terms $T_i, \; i > -s.$ 
Since the algebra $ (C^{\infty}[\lambda^{-1},\lambda]]({\cal M}), \bullet)$ is in fact classical, this result is logical. But in quantum considerations eigenstates are expected to have supports of nonzero measures.
As one can see, Eqs. (\ref{3.61}) do not determine the number $s.$
 
A reasonable remedy for this problem with supports would be  an extension of  the formal calculus presented before on the Laurent series of
generalised functions 
\[
 \sum_{k=1}^{\infty} \lambda^{-k}T_{-k} + \sum_{k=0}^{\infty} \lambda^{k}T_{k}. 
 \]
  This suggestion is of a deep
physical origin. Indeed,  Wigner functions  on the phase space ${\mathbb R}^{2r}$ in their Laurent expansions contain arbitrary negative powers of  the Planck constant (see Sec. \ref{sec4}). Thus we are going to consider that extension of ${\cal E}'({\cal M}).$

Unfortunately, we see that the action
\be 
\label{6} 
\Big< \sum_{k=1}^{\infty} \lambda^{-k}T_{-k} + \sum_{k=0}^{\infty} \lambda^{k}T_{k},
\sum_{m=0}^{\infty} \lambda^{m} \varphi_{m} \Big>= \sum_{z=1}^{\infty} \lambda^{-z} \sum_{k=0}^{\infty}\Big< T_{-z-k},\varphi_k \Big>
+ \sum_{z=0}^{\infty} \lambda^{z} \sum_{k=0}^{\infty}\Big< T_{z-k},\varphi_k \Big>
 \ee 
 is not well defined.
 If the part $\sum_{k=1}^{\infty}
\lambda^{-k}T_{-k}$ of the Laurent series has got infinitely many elements, the sums $\sum_{k=0}^{\infty}\Big< T_{-z-k},\varphi_k \Big>$ and $\sum_{k=0}^{\infty}\Big< T_{z-k},\varphi_k \Big>$ may be divergent.
 Generalisation of this conclusion on   formal series of the kind $\sum_{m=-u}^{\infty} \lambda^{m} \varphi_{m}$  is straightforward.
 
 Therefore we can see that formal series $\sum_{k=1}^{\infty} \lambda^{-k}T_{-k} + \sum_{k=0}^{\infty} \lambda^{k}T_{k}$ of generalised functions belonging to  ${\cal E}'(\cal{M})$ are not functionals over the space of formal series $ C^{\infty}[\lambda^{-1},\lambda]]({\cal M}).$


\subsection{States over the algebra $ (C^{\infty}[\lambda^{-1},\lambda]]({\cal M}), *)$ }
\label{sub3.3}

Considerations devoted to the states over the nontrivially deformed algebra $ (C^{\infty}[\lambda^{-1},\lambda]]({\cal M}), *)$ ought to start from an essential reformulation of the
action of a linear functional. Indeed, since both: the pointwise $\cdot$ -- multiplication of functions  and the $\bullet$--product were Abelian, the functional calculus 
 originating on the integral (\ref{03}) satisfies the property of traciality.

Now the situation is more complicated. If a function $ \psi \in (C^{\infty}({\cal M}), *)$ is of a compact support and does not contain the
deformation parameter $\lambda,$ then its action on an arbitrary function $\varphi \in C^{\infty}({\cal M})$ should be of the form
\be
\label{3.001}
{\mathbb C}[\lambda^{-1}, \lambda]] \, \ni \,
\big< \psi, \varphi \big>_{*} := \frac{1}{\lambda^r} \int_{\cal M}(\psi * \varphi) \bullet t[[\lambda]]\, \omega^r.
\ee
In general of course $\big< \psi, \varphi \big>_{*} \neq \big< \psi, \varphi \big>,$ which means that the same function $\psi$ determines different functionals depending on the choice of the $*$--product in the set of functions $C^{\infty}({\cal M}).$ Application of the idea of functional action (\ref{3.001}) to  formal series $\psi[[\lambda]], \varphi[[\lambda]] \in C^{\infty}[\lambda^{-1},\lambda]]({\cal M})$ is natural.
The functional action (\ref{3.001}) satisfies the requirement of traciality. 

  ${\mathbb C}[\lambda^{-1}, \lambda]]$--linearity of the functional  (\ref{3.001}) is obvious. Moreover, if the $*$ -- product is continuous, the formula (\ref{3.001}) defines a continuous functional over the space $ C^{\infty}({\cal M}).$ Therefore one can notice that {\it the functional  calculus based on  formula (\ref{3.001}) is equivalent to the standard theory of generalised functions} with an identification
 \[
 \psi \sim T_{\psi}[[\lambda]]:=\frac{1}{\lambda^r} t[[\lambda]] \bullet \sum_{l=0}^{\infty} \lambda^l T_{\psi \, l} \in {\cal E}' [\lambda^{-1}, \lambda]]({\cal M})\;\;\; {\rm i.e.}
 \]
 \be
 \label{3.002}
 \forall \, \varphi \in  C^{\infty}({\cal M}) \;\; \big< \psi, \varphi \big>_* = \big< T_{\psi}[[\lambda]], \varphi \big>.
 \ee
 We sum from $l=0$ because the $*$--product introduces only terms standing at nonnegative powers of $\lambda.$ The relation between the function $\psi$ and its respective formal functional  $T_{\psi}[[\lambda]]$ depends on the $*$ -- multiplication. The value of the right hand side of (\ref{3.002}) is calculated exactly as in the case of trivial deformation. Since the $*$ --product is local, for every index $l$  the relation ${\rm supp} \,T_{\psi \,l}  \subset {\rm supp} \,\psi$ is satisfied.
 
 The natural question arises, if two different functions $\psi,\psi'$ may lead to the same formal  generalised function $T_{\psi}[[\lambda]]=T_{\psi'}[[\lambda]]$ in (\ref{3.002}). 
 Applying the definition of the $*$--product one gets that the term of the order $\lambda^{-r}$ in (\ref{3.001}) is 
 \[
\int_{\cal M} \psi \cdot \varphi \, t_0 \, \omega^r,
 \]
   as the  trace  density $ t[[\lambda]]= \sum_{l=0}^{\infty} \lambda^l t_l$.
 But this integral expresses the functional action for a generalised function determined by the  function $\psi \,t_0.$ To recover the classical limit one  puts $t_0=1.$ In this limit we can see that since the functions $\psi$ and $\psi'$ are smooth, they determine different functionals.
 
 The observation (\ref{3.002}) indicates the way in which we can adapt the integral action (\ref{3.001}) for an arbitrary element of the space ${\cal E}' [\lambda^{-1}, \lambda]]({\cal M}).$ 
 Every formal  series of generalised functions $T[[\lambda]]= \sum_{l=-s}^{\infty}\lambda^l T_l$ is the limit of  a series of functionals defined by formal series of smooth functions of compact supports
 $
 T[[\lambda]]= \lim_{n \rightarrow \infty} (\sum_{l=-s}^{\infty} \lambda^l \psi_{l})_n
 $
 in the sense of convergence (\ref{003}). Therefore by definition
 \be
 \label{3.003}
 \forall\;\; \varphi\, \in  C^{\infty}({\cal M})  \,\;\;\;
 \big< T[[\lambda]], \varphi \big>_* := \lim_{n \rightarrow \infty} \big< (\sum_{l=-s}^{\infty} \lambda^l \psi_{l})_n, \varphi \big>_*.
 \ee
An extension of this result on an arbitrary formal series $ \varphi[[\lambda]] \in  C^{\infty}[\lambda^{-1},\lambda]]({\cal M})$ is straightforward. 

Quantum states should fulfill the requirements analogous to these ones satisfied by  states over the trivial algebra $(C^{\infty}[\lambda^{-1},\lambda]]({\cal M}), \bullet).$ Therefore  we assume that
 quantum states are the  linear continuous functionals over the space of formal series 
 $C^{\infty}[\lambda^{-1},\lambda]]({\cal M}),$
 which are real, nonnegative and  satisfy the normalisation condition. The notion of reality means that 
\[
\forall \varphi[[\lambda]] \in C^{\infty}[\lambda^{-1},\lambda]]({\cal M})\;\; {\rm such\;\; that }\;\;
\overline{\varphi[[\lambda]]}= \varphi[[\lambda]]\;\;\;{\rm there \;\; is}
\] 
\be
 \label{3.21q}
   \big< T[[\lambda]], \varphi[[\lambda]] \big>_* = \overline{\big< T[[\lambda]], \varphi[[\lambda]] \big>_*}.
\ee
Thus the requirement of reality for a series of generalised functions $T[[\lambda]]= \sum_{l=-s}^{\infty} \lambda^l T_l$ says in  fact that for every index `$l$' and every real function $\varphi \in C^{\infty}({\cal M})$ there is
\be
\label{q1}
\big<T_l, \varphi \big>_* = \overline{\big<T_l, \varphi \big>_*}.
\ee
In order to introduce nonnegativity we will follow the route proposed for the trivially deformed algebra  
 $(C^{\infty}[\lambda^{-1},\lambda]]({\cal M}), \bullet).$
 \begin{de} 
A formal series $\sum_{l=-s}^{\infty} \lambda^l T_l \in {\cal E}'[\lambda^{-1},\lambda]]({\cal M})$ is nonnegatively defined if for every series $ \sum_{m=-u}^{\infty} \lambda^{m} \varphi_{m} \in C^{\infty}[\lambda^{-1},\lambda]]({\cal M})$
the formal series of real numbers
\be 
\label{5.50}
 \Big< \sum_{l=-s}^{\infty} \lambda^l T_l, \sum_{m_1=-u}^{\infty}
\lambda^{m_1} \overline{\varphi}_{m_1} * \sum_{m_2=-u}^{\infty} \lambda^{m_2} \varphi_{m_2}\Big>_* 
 \ee 
 is nonnegative in the sense of Def. \ref{pos4}.
\end{de}
An alternative approach is the following. 
  Let us  introduce a formal series ${\cal T}_T[[\lambda]]$ generalising the identification rule
(\ref{3.002})
\be
\label{q1.1}
\forall\; \varphi \in C^{\infty}({\cal M})\;\;
\big< {\cal T}_T[[\lambda]], \varphi \big>:=
\big< T[[\lambda]], \varphi \big>_*.
\ee
It is of the form  $ {\cal T}_T[[\lambda]]= \sum_{l=-u}^{\infty} \lambda^l {\cal T}_l.$ By analogy to the cases considered before   we see that this functional  ought to act on the  product of formal series $\overline{\varphi[[\lambda]]}* \varphi[[\lambda]].$ Such product is  real   because the $*$--product is Hermitian. Therefore
 $\overline{\overline{\varphi[[\lambda]]}* \varphi[[\lambda]]}=\overline{\varphi[[\lambda]]}* \varphi[[\lambda]]$ 
 for every $ \varphi[[\lambda]] \in C^{\infty}[\lambda^{-1},\lambda]]({\cal M}).$
 
According to  Corollary 
\ref{cowazny} we  consider only functions from  
the set $C^{\infty}({\cal M})$ and
 formulate the following definition of nonnegativity.
\begin{de}
A formal series $T[[\lambda]] \in  {\cal E}' [\lambda^{-1}, \lambda]]({\cal M})$ represented by the series ${\cal T}_T[[\lambda]]= \sum_{l=-u}^{\infty} \lambda^l {\cal T}_l$ is nonnegatively defined, if for every value $\lambda>0$ and every function $\varphi \in C^{\infty}({\cal M}) $  there exist natural  numbers $u$ and $m$ such that 
\be
\forall\; k>u \; \forall \;r>m \;\;\;\Big< \sum_{l=-s}^k \lambda^l {\cal T}_l, \sum_{z=0}^r \, \lambda^z B_z(\overline{\varphi},\varphi) \Big> \geq 0.
\ee
\end{de}
As we will show in an example, the property of nonnegativity is the most difficult to be satisfied.

Finally, the normalisation condition states that
\be 
\label{3.4q} 
\big<T[[\lambda]], {\bf 1} \big>_*=1.
 \ee 
Since every formal series $T[[\lambda]]$ is a limit of a sequence of formal series of smooth functions with compact supports, the requirement (\ref{3.4q}) is equivalent  to the set of conditions
\be
\label{q3}
\frac{1}{\lambda^r}
\big<T_r,{\bf 1} \big>=1\;\;\; {\rm and}\;\;\;\big<T_k,{\bf 1} \big>=0
 \;\;\;{\rm for}\;\;\;  k \neq r.
 \ee
Notice that in the normalisation condition  is sufficient to use the standard functional action $\big<T_r,{\bf 1} \big>$ instead of $\big<T_r,{\bf 1} \big>_*.$
\vspace{0.3cm}

Let $\xi[[\lambda]] \in C^{\infty}[\lambda^{-1},\lambda]]({\cal M}) $ represent an observable. Its eigenstate assigned to an eigenvalue $a[[\lambda]] \in {\mathbb R}[\lambda^{-1},\lambda]]$ is a state $T_a[[\lambda]] \in {\cal E}' [\lambda^{-1}, \lambda]]({\cal M})$ fulfilling the system of equations 
\setcounter{orange}{1} \renewcommand{\theequation}{\arabic{section}.\arabic{equation}\theorange}
\be
\label{q41}
\xi[[\lambda]] * T_a[[\lambda]] = a[[\lambda]] \cdot T_a[[\lambda]] \;\;\; {\rm and}
\ee
 \addtocounter{orange}{1}
\addtocounter{equation}{-1} 
\be
\label{q42}
\xi[[\lambda]] * T_a[[\lambda]]= T_a[[\lambda]]* \xi[[\lambda]].
\ee
\renewcommand{\theequation}{\arabic{section}.\arabic{equation}} 
Since functionals are defined by their action on test functions, the requirement (\ref{q41}) means that
\setcounter{orange}{1} \renewcommand{\theequation}{\arabic{section}.\arabic{equation}\theorange}
\be
\label{q51}
\forall\; \varphi \in C^{\infty}({\cal M}) \;
\big< (\xi[[\lambda]]-a[[\lambda]]) * T_a[[\lambda]], \varphi \big>_*=0
\ee
 \addtocounter{orange}{1}
\addtocounter{equation}{-1} 
or equivalently
\be
\label{q52}
\forall\; \varphi \in C^{\infty}({\cal M}) \;
\big< (\xi[[\lambda]]-a[[\lambda]]) * T_a[[\lambda]], \varphi \big>=0.
\ee
However we have not defined  yet the products $\xi[[\lambda]] * T_a[[\lambda]]$ and $T_a[[\lambda]]* \xi[[\lambda]].$
By analogy to the pointwise product of a generalized function by a smooth function we introduce these multiplications by the following conditions
\setcounter{orange}{1} \renewcommand{\theequation}{\arabic{section}.\arabic{equation}\theorange}
\be
\label{q61}
\forall\; \varphi, \xi \in C^{\infty}({\cal M}) \;\;\;
\big<\xi* T, \varphi \big>_*:=  \big<T, \varphi * \xi \big>_*
\ee
 \addtocounter{orange}{1}
\addtocounter{equation}{-1}
and 
\be
\label{q62}
\forall\; \varphi, \xi \in C^{\infty}({\cal M}) \;\;\;
\big<T* \xi, \varphi \big>_*:=  \big<T, \xi * \varphi  \big>_*
\ee
\renewcommand{\theequation}{\arabic{section}.\arabic{equation}} 
Generalisations of formulas (\ref{q61}), (\ref{q62}) on formal series are straightforward. 

Assume that $\xi[[\lambda]]=\sum_{i=0}^{\infty} \lambda^i \xi_i$ because the multiplication by an arbitrary power of the parameter  $\lambda$ (also negative) does not influence  the eigenvalue equation (\ref{q41}). Since  the $*$ -- product generates only terms standing at nonnegative powers of $\lambda$, one can consider only generalised eigenfunctions of the form  $T_a[[\lambda]]= \sum_{i=0}^{\infty} \lambda^i T_{a \,i}.$ Thus there must be $a[[\lambda]]=\sum_{i=0}^{\infty} \lambda^i a_i.$
 Then the eigenvalue equation (\ref{q41}) separates into an infinite system of equations
\bea
\label{q7}
(\xi_0-a_0)T_0 & = & 0, \nonumber \\
(\xi_1-a_1)T_0 + (\xi_0-a_0)T_1 +B_1 \big( \xi_0-a_0,T_0 \big) & = & 0 , \\
\vdots & & \vdots \nonumber
\eea
The product $B_1 \big( \xi_0-a_0,T_0 \big)$ can be calculated in the following way. We construct a series $\{u_n\}_{n=1}^{\infty}$ of smooth functions of a common compact support, which is convergent to the generalised function $T_0.$ Then for every $k\geq 0 $
\be
\label{q8}
B_k(\xi,T_0):=  \lim_{n\rightarrow \infty}B_k(\xi, u_n).
\ee
Since the $*$--product is local, of course ${\rm supp }\,B_k(\xi,T_0) \subset {\rm supp} \,\xi \cap {\rm supp}\, T_0. $ 

From the first equation of the system (\ref{q7}) we obtain that
$
{\rm supp}\;T_0 
$ is contained in the area fulfilling the condition $\xi_0-a_0=0,$ which can be of measure zero.
The second equation  implies that the support of $T_1$ is not greater than  the support of $T_0.$ Following this way we arrive to  hardly acceptable conclusion that in many cases  supports of quantum eigenstates are of measure zero.

  We have been still investigating this option  however at this moment  
 we propose extension of  the set of states by  Laurent  series of generalised functions in  $\lambda$ with infinite principal part. This idea also causes formidable obstacles. As we know from the subsection devoted to the trivially deformed algebra $(C^{\infty}[\lambda^{-1},\lambda]]({\cal M}), \bullet),$ the action of formal  Laurent series on test functions is in general  not well defined.  Thus the same happens for the algebra  $(C^{\infty}[\lambda^{-1},\lambda]]({\cal M}), *),$
which is a deformation of $(C^{\infty}({\cal M})[\lambda^{-1},\lambda]], \bullet).$
\section{Example of the Moyal algebra}
\label{sec4}
\setcounter{equation}{0}

To illustrate differences between states over the trivial algebra $(C^{\infty}[\lambda^{-1},\lambda]]({\cal M}), \bullet)$ and the nontrivially deformed one $(C^{\infty}[\lambda^{-1},\lambda]]({\cal M}), *)$ let us consider the best known $*$  -- algebra namely the algebra $(C^{\infty}[\lambda^{-1},\lambda]]({\mathbb R}^2), *_{ M})$ with the Moyal product.
By definition \cite{pleban, ja1} this product is calculated as
\[
\forall \;\; \varphi, \psi \in C^{\infty}({\mathbb R}^2)
\]
\be
\label{q31}
\varphi *_{M} \psi := \sum_{n_1, n_2=0}^{\infty} \frac{1}{n_1 ! n_2 !} \left(- \frac{i \lambda}{2}\right)^{n_1}
\left( \frac{i \lambda}{2}\right)^{n_2} \frac{\partial^{n_1+n_2} \varphi}{\partial p^{n_1} \partial q^{n_2}}
 \frac{\partial^{n_1+n_2} \psi}{\partial q^{n_1} \partial p^{n_2}}
\ee
and this way of multiplication can be easily extended on formal series belonging to the set $C^{\infty}[\lambda^{-1},\lambda]]({\mathbb R}^2).$

The Moyal product is closed i.e. if at least one of the multiplied function is of a compact support then
\be
\label{q32}
\int_{{\mathbb R}^2} \varphi *_{M} \psi \, dq dp= \int_{{\mathbb R}^2} \varphi \cdot \psi \, dq dp.
\ee
Therefore an identification holds
\[
\forall\, \varphi[[\lambda]] \in C^{\infty}[\lambda^{-1}, \lambda]]({\mathbb R}^2)\;\; {\rm and} \;\; \forall \,T[[\lambda]] \in {\cal E}' [\lambda^{-1}, \lambda]]({\mathbb R}^2)
\]
\be
\label{q33}
\Big<T[[\lambda]], \varphi[[\lambda]] \Big>_{*_M} = \frac{1}{\lambda}\Big<T[[\lambda]], \varphi[[\lambda]] \Big>.
\ee
Thus reality of the functional $T[[\lambda]]= \sum_{l=-s}^{\infty}\lambda^l T_l$ means that  every generalised 
function $T_l, \; l=-s, -s+1, \ldots $ is real.

Essential obstacles reveal when we consider nonnegativity.
On the contrary to the case of $\bullet$--multiplication the product $\overline{\varphi} * \varphi$ may be locally negative. Indeed, e.g.  function 
\[
\Big( (q-{\tt q}_0)-ia(p-{\tt p}_0) \Big)*\Big( (q-{\tt q}_0)+ia(p-{\tt p}_0) \Big), \;\;\; {\tt q}_0, {\tt p}_0,a \in {\mathbb R}, \; a>0,
\]
  is negative inside the ellipse
\[
\frac{(q-{\tt q}_0)^2}{a \lambda} + \frac{(p-{\tt p}_0)^2}{ \lambda / a}=1.
\]
The area of this ellipse equals $\pi \lambda.$ 
Thus any classical state $T \in {\cal E}'({\mathbb{R}^2})$ represented by a summable function of compact support, can be excluded from being quantum state, by appropriate scaling of $\lambda$.

Moreover, we claim that any series $T[[\lambda]]]= \sum_{l=-s}^{\infty}\lambda^l T_l$, where $T_l$ are summable functions of supports contained in a single common compact set ${\rm supp}( T[[\lambda]])$, cannot represent a quantum state. This statement is supported by the following observation. By WKB approximation a volume occupied by a single quantum state is proportional to $\lambda$. The same conclusion can be reached from the Heisenberg uncertainty principle.  Simultaneously our definition of nonnegativity involves every $\lambda > 0$, so one can make this volume arbitrarily large. On the contrary, the volume defined by ${\rm supp}\; T[[\lambda]]$ does not depend on $\lambda$.

These observations are related to another aspect of the formal series calculus. The deformation parameter $\lambda$ has its own unit. If we fix its value, then an arbitrary change of this unit does not change  convergence (divergence) of a series $\varphi[[\lambda]]$ because the units of variables $(p,q)$ also change and in turn the same happens to the units of coefficients of $\varphi[[\lambda]]$. But when $\lambda$ is treated as a symbol, we lose this invariance of convergence (divergence) with respect to the value of $\lambda.$

By such arguments one is driven to represent states in frames of the Moyal formal series calculus  by generalised functions of noncompact supports. What is exciting, this observation is with agreement with a result holding for the Moyal multiplication in strict deformation. For the phase space $({\mathbb R}^2, dq \wedge dp)$ equipped with the Moyal product (\ref{q31}) a relationship between a wave function and its respective state called a Wigner function, is known \cite{ja1}. And as it was proved in \cite{ja2}, if the wave function is of a compact support then its  Wigner function is not. One can extend this observation on mixed states.

Indeed, let us take a look at Wigner eigenfunctions of the harmonic oscillator. As it was shown \cite{GW46, taka} they are of the form
\be
\label{wp1}
W_n(H(q,p))= \frac{(-1)^n}{\pi \hbar } \exp \left( - \frac{2H(q,p)}{\hbar \omega}\right)L_n\left( \frac{4H(q,p)}{\hbar \omega}\right),
\ee
where the index $n=0,1,\ldots$  numbers  energy levels and $H(q,p)= \frac{p^2}{2m} + \frac{m \omega^2 q^2}{2}$ is the Hamilton function. The deformation parameter $\lambda$ is represented by the Planck constant $\hbar$. By $L_n$ the Laguerre polynomials are denoted.

  By direct calculations one can check that the Wigner function (\ref{wp1}) is a formal Laurent series with respect to $\hbar$
  \be
  \label{wp2}
  W_n(H(q,p))= \frac{(-1)^n}{\pi \hbar } \sum_{z=0}^{\infty} \frac{1}{\hbar^z} \frac{1}{z!} \left( - \frac{2 H(q,p)}{\omega}\right)^z \times 
  \sum_{m=0}^{{\rm min.}[z,n]} 2^m \left( \begin{array}{c}n \\m \end{array}\right) 
  \left( \begin{array}{c}z \\m \end{array}\right).
  \ee
  Thus we see that the greatest power of the deformation parameter is $0$ and that all terms standing at the series (\ref{wp2})  are of noncompact support. Therefore the mean value of energy equals
  \be
  \label{wp3}
  \big< H(q,p)\big>= \big<W_n(H(q,p)), H(q,p) \big>= \frac{1}{\omega}\int_{0}^{2\pi}d \phi \int_{0}^{\infty}d H \,W_n(H) H= \hbar \omega \left(n + \frac{1}{2} \right)
  \ee
  where we express the Wigner function by (\ref{wp1}) and apply the change of variables $(q,p) \rightarrow (H, \phi).$  On the other hand the integral (\ref{wp3}) is not defined for  $W_n(H)$ represented by the series (\ref{wp2}). Factor $\frac{1}{\omega}$ standing in front of the integral results from the change of variables. 
  


\section{Conclusions}
In our paper we discuss a few possible extensions of the theory of generalised functions over the space of formal series of smooth functions. Constructions of those extensions are natural. 

Problems appear when one tries to introduce nonnegativity of the proposed formal series. The property of nonnegativity is crucial for possible physical applications because states are represented by such functionals. There are two main causes of the difficulties. One obstacle  originates from the fact that on the one hand we do not refer to any fixed value of the deformation parameter and on the other hand we ask whether the result of functional action is nonnegative. The second one is related to the $*$-- product. As we showed, in some algebras
$ (C^{\infty}[\lambda^{-1},\lambda]]({\cal M}), *)$  the product $\overline{\varphi} * \varphi$ may be locally negative. In turn there is clear tension between quantum mechanical arguments and the postulate of compactness of supports of states.
 

Another challenge are eigenstates of quantum observables. They are usually represented by formal series of generalised functions containing arbitrary negative powers of the deformation parameter. And such series are not distributions over the set of smooth functions.

Thus we conclude sadly that in general it is hardly possible to put the formal series calculus adequate for  quantum mechanics in frames of the mathematically rigorous theory of generalised functions. But it does not mean that one cannot apply formal series in quantum problems at all. Just one needs to treat  observables individually since every measurable quantity  has its own class of admissible states. This method is not as elegant as dealing with one universal class of test functions and  its unique set of linear functionals but fully acceptable. Quite surprisingly some hope arises when one considers states described by generalised functions of zero measure supports. This kind of analysis is going to be covered by our forthcoming article. 


A remedy for the aforementioned difficulties could be a strict quantisation \cite{berezin}. The strict quantisation  would eliminate formal series and produce convergent expressions. An extra profit from this convergence would be possibility of substituting a numerical value of the deformation parameter. Unfortunately at this moment the strict quantisation procedure is known only for some specific types of manifolds, see e.g.  \cite{gra}.

\end{document}